# THE SECOND BYURAKAN SURVEY GALAXIES. I. THE OPTICAL DATABASE

**M. GYULZADYAN[1], B. McLEAN[2], V.Zh.ADIBEKYAN[1,3], R. J. ALLEN[2], D. KUNTH[4], A. PETROSIAN[1], J. A. STEPANIAN[5]**

*A database for the entire catalog of the Second Byurakan Survey (SBS) galaxies is presented. It contains new measurements of their optical parameters and additional information taken from the literature and other databases. The measurements were made using $I_{pg}$(near-infrared), $F_{pg}$(red) and $J_{pg}$(blue) band images from photographic sky survey plates obtained by the Palomar Schmidt telescope and extracted from the STScI Digital Sky Survey (DSS). The database provides accurate coordinates, morphological type, spectral and activity classes, apparent magnitudes and diameters, axial ratios, and position angles, as well as number counts of neighboring objects in a circle of radius 50 kpc. The total number of individual SBS objects in the database is now 1676. The 188 Markarian galaxies which were re-discovered by SBS are not included in this database. We also include redshifts that are now available for 1576 SBS objects, as well as 2MASS infrared magnitudes for 1117 SBS galaxies.*

## 1. Introduction.

About a half a century ago, V. Ambartsumian's concept of nuclear activity [1] and the discovery of the wide variety of active and starburst galaxies created new directions of research in astronomy. In particular, it encouraged the study of nuclear activity and the star formation history of galaxies. One of the crucial ways to increase our knowledge in this field is the detailed study of large numbers of active and star-forming galaxies discovered via surveys in different wavelengths.

---

[1] V. A. Ambartsumian Byurakan Astrophysical Observatory,
  Armenia, e-mail: mgyulz@bao.sci.am, e-mail: artptrs@yahoo.com,
[2] Space Telescope Science Institute, Baltimore, USA, e-mail: mclean@stsci.edu,
[3] Yerevan State University, Armenia, e-mail: adibekyan@bao.sci.am,
[4] Institut d' Astrophysique de Paris, Paris, France
[5] 3D Astronomy Center, Mexico

Since the pioneering studies of Haro [2], Zwicky [3, 4] and Markarian [5], many optical surveys have been carried out using the wide-field imaging capabilities of Schmidt telescopes with special detection techniques based on some tracer of galaxy activity (e.g. enhanced blue color, UV-excess radiation or existence of emission lines).

The technique of color selection was introduced by Haro [2], and was used in the first and second Kiso surveys [6,7] as well as the ongoing Montreal survey [e.g. 8]. The technique of selecting objects according to their excess UV emission on low-dispersion objective-prism spectra was introduced by Markarian [3]. Objects in the Markarian survey were selected on the criterion that the blue-violet portion of the spectrum is brighter and more extended in wavelength than the red-yellow portion. For A0-A2 stars, these two portions (separated by the "green dip") have approximately similar brightness and extension. This survey produced a catalog of 1515 active and star-forming galaxies [9]. Another survey which used the same telescope and methods of observation but covering different sky area was conducted by Kazarian [10]. The most popular method for discovering active and star-forming galaxies is searches for the presence of emission-lines in low-dispersion, objective-prism spectra. These include the CTIO [e.g. 11], UM [e.g. 12], Wasilewski [13], UCM [e.g. 14], HQS [e.g. 15] and Hamburg/SAO [e.g. 16] surveys. Recently, the benefits of objective-prism observations with Schmidt telescopes have been combined with CCD detectors to provide the means for sensitive new emission-line surveys (e.g. KISS [17]).

The combination of both UV excess and emission lines methods allows the discovery of objects with a broader range of star formation histories and with a larger variety of activity [18]. Surveys that use this combined technique to discover active and star-forming galaxies include the Case survey [e.g. 19], the Second Byurakan Survey [e.g. 20], and the Marseille Schmidt survey [21]. In recent years, the availability of wide field surveys at many wavelengths from radio to X-ray [e.g. 22-25] as well as the Sloan Digital Sky Survey (SDSS) [26], has resulted in the discovery of large number of active and star-forming galaxies [e.g. 27,28]. In order to extract reliable and useful information from these catalogs, they must be as complete and representative as possible with the purpose of controlling selection effects and also discussing in much detail the nature of the objects included in the catalog.

In this paper we describe the properties of the Second Byurakan Survey (SBS) galaxies, which are identified because of their UV excess emission and/or presence of emission lines. The approach which was used to prepare this paper is very similar to that which was used in the original paper of Markarian galaxies [29]. This paper describes a new database of 1676 SBS galaxies containing the following new measurements: accurate optical positions, morphological classes, apparent near-infrared, red and blue magnitudes, diameters, axial ratios and position angles, and finally counts of neighbor galaxies within a 50 kpc circle radius, based on the galaxy redshift and assuming a value for the Hubble constant of $H_0 = 75$ km s$^{-1}$ Mpc$^{-1}$. In addition, we have compiled from the literature updated spectral and activity classes, as well as new and revised determinations of redshifts. In the database, we have also included the corresponding infrared magnitudes for all the galaxies from 2MASS survey. This will facilitate comparison of the optical and near-infrared properties of galaxy subsets.

In Section 2 of this paper, we describe the SBS survey and some results. It also describes the observational material and the generation of the database. The database itself is described in Section 3. The next paper in this series will present multi-band photometric and spectrophotometric data for all SBS galaxies which were observed by SDSS.

## 2. The Second Byurakan Survey.

The SBS is a continuation of the Markarian survey [5] with the goal of reaching fainter objects and discovering new active and star-forming galaxies using both UV excess and emission line techniques.

Both the Markarian survey and the SBS have been carried out with the 40 – 52 inch Schmidt telescope of the Byurakan Observatory in Armenia, but for the SBS the telescope was equipped with three objective prisms (1º.5, 3º, and 4º) instead of only one (1º.5) for Markarian survey. In addition, more sensitive, "baked" Kodak IIIaJ, IIIaF, and IV-N photographic plates were used, reaching a limiting photographic magnitude of about 19.5, which is about 2.5 mag fainter than the Markarian survey limit. In addition to discovering peculiar objects with strong UV-excess radiation, the improved spectral resolution of the wider angle prisms permitted to identify of galaxies with moderate and strong emission lines even if the UV-excess emission was absent. Objects with observed peculiar energy distribution were also selected. The SBS objective-prism observations started in 1974 and finished in 1986. SBS plates cover the region of sky defined by $7^h40^m < \alpha < 17^h15^m$, $49° < \delta < 61°$. The survey consists of 64 fields (635 plates) 4° x 4° in size and covered a total area of 991 square degrees.

The first list of the SBS objects was published in 1983 [30] and was followed by six more lists [31–36]. The results of the SBS survey are summarized in the SBS general catalog [37], which presents detailed information about the SBS survey area and its structure, observational strategy, object selection methods and their classification criteria. The general characteristics of the survey are summarized in Table 3 of this paper. The SBS general catalog [37] consists of 3563 objects presented in two parts: a catalog of galaxies (1863 objects) and a catalog of stellar objects (1700 objects). The former SBS catalog included 761 new AGNs. The SBS sample is found to be complete at 70% for galaxies and about 85% for AGNs/QSOs with $B \leq 17.5$ [37].

Table 1 - Database of SBS Galaxies

| SBS | RA | Dec | Morph | SC | AC | z | Jpg | Fpg | Ipg | D"(J) | R(J) | PA(J) | N | J | H | K |
|---|---|---|---|---|---|---|---|---|---|---|---|---|---|---|---|---|
| (1) | (2) | (3) | (4) | (5) | (6) | (7) | (8) | (9) | (10) | (11) | (12) | (13) | (14) | (15) | (16) | (17) |
| 0742+599 | 7 46 34.63 | + 59 51 28.2 15 | 15 | d2e | e | 0.0328 | 17 | 16 | 16 | 29.58 | 0.4 | 111 | 1 | 14.55 | 13.79 | 13.63 |
| 0743+591 A | 7 47 21.18 | + 59 1 7.2 3 | 3 | sde | e,a | 0.0325 | 14 | 14 | 13 | 63.24 | 0.4 | 75.9 | 0 | 11.7 | 10.9 | 10.39 |
| 0743+591 B | 7 47 45.95 | + 59 0 27.7 15 | 15 | se | | 0.0211 | 18 | 17 | 17 | 9.18 | 1 | | 1 | | | |
| 0743+591 C | 7 47 58.72 | + 59 0 52.8 4 | 4 B | se: | e,a | 0.0217 | 14 | 14 | 13 | 110.2 | 0.5 | 17.6 | 1 | 11.9 | 11.18 | 11.05 |
| 0744+502 | 7 48 6.97 | + 50 6 49.5 7 | 7 | dse | | 0.0215 | 17 | 16 | 16 | 45.9 | 0.9 | 112 | 0 | 14.65 | 14.15 | 13.93 |
| 0745+557 | 7 49 10.05 | + 55 36 16.7 3 | 3 B | s3e | SB | 0.0169 | 15 | 14 | 14 | 49.98 | 0.4 | 75.7 | 0 | 12.28 | 11.55 | 11.3 |
| 0745+571 | 7 49 10.51 | + 57 2 51.5 -1 | -1 | dse | e | 0.0439 | 15 | 14 | 14 | 40.8 | 0.9 | 36.5 | 0 | 12.3 | 11.38 | 11.07 |
| 0745+587 | 7 49 36.57 | + 58 39 37.4 15 | 15 | sd3e | | 0.0212 | 16 | 15 | 15 | 21.42 | 0.4 | 146 | 0 | | | |
| 0745+590 | 7 49 16.85 | + 58 55 11 15 | 15 | d2e | HII | 0.0274 | 16 | 15 | 15 | 28.56 | 0.5 | 112 | 0 | 13.32 | 12.77 | 12.46 |
| 0745+598 | 7 50 14.08 | + 59 41 14.7 0 | 0 | dse | e,a | 0.0700 | 19 | 18 | 17 | 18.36 | 0.3 | 2.2 | 0 | | | |

The subject of this study is 1676 SBS galaxies. These objects are selected from the total sample of 1863 SBS galaxies excluding 188 Markarian galaxies (see Table 7 of [37]). These 188 galaxies being included in total sample of 1544 Markarian galaxies were studied in [29]. We added one object, SBS1204+554A, with interesting spectra in SDSS DR7 to the database ("A" designation is added by us). This object was included in the original list of SBS [33] but is not present in the SBS general catalog [37].

The SBS General Catalogue [37] presents a large number parameters for SBS galaxies, but there still remains a significant incompleteness in many fundamental parameters such as magnitudes, diameters and morphologies. Today the availability of the high-quality observations from sky surveys allows us to measure additional optical parameters such as morphology, apparent magnitude, size, and axial ratio of galaxies in a more accurate and homogeneous way, and also to extract quantitative data related to their local environment. The data gathered in this database were measured and compiled with the goal of obtaining a complete, homogeneous set of optical parameters for SBS galaxies for further statistical studies.

The Sloan Digital Sky Survey covers approximately the area observed by the SBS. In the currently available Data Release 7 (DR7) of SDSS [38] data for 1540 SBS galaxies are available, but it should be noted that in this paper we do not include any data from SDSS except corrected redshifts. We plan to present SDSS photometric as well as spectrophotometric data and several derived parameters for observed SBS galaxies in the next article of this series.

## 3. Compiled parameters for SBS galaxies.

Some of the parameters presented for the SBS galaxies in this database are compiled from different sources. These parameters are the following:

**3.1. Redshifts.** In this database heliocentric redshifts of 1576 out of 1676 SBS galaxies are taken mostly from [37], from the NASA extragalactic database (NED) and from the SDSS. In cases when more than one redshift measurements are available, the more accurate 21 cm or the latest published value (usually from DR7 of SDSS) is given. All the remaining 100 objects without redshift are isolated galaxies. In [37] is stressed that for several SBS galaxies two published values cardinally differ from each other. We also add to this list the following objects: SBS1146+596, z([37])=0.0107, z(DR7)=1.9468, SBS1323+600W, z([37])=0.0374, z(DR7)=0 (it is a galactic star), SBS1433+554S, z([37])=0.0730, z(DR7)=0.1403.

**3.2. Spectral classification**. In the database, spectral classifications for all SBS galaxies are listed. This spectral classification system was originally introduced by Markarian [5], for the First Byurakan Survey objects. It describes the degree of spatial concentration of the UV emission as well as its intensity. Emission regions are classified as stellar ("s") or diffuse ("d") if the half-width of the emission region on Schmidt plates is about 2" or 6" – 8", respectively. The intermediate types "sd" and "ds" were also used. A number between 1 and 3 was used to indicate the relative intensity of the UV emission, with 1 being the strongest UV excess. The existence of emission lines in the spectra was marked with "e" or with "e:" for uncertain cases.

In the SBS lists, objects with excess UV radiation have spectral classifications similar to that of Markarian [5]. Objects which have no UV excess but with moderate or strong emission lines have been classified according to the degree of concentration of their continuum radiation. The paper [39] compares SBS galaxies discovered using UV excess with those found only from their strong emission lines. According to this study there are several interesting differences in both populations of SBS objects.

**3.3. Activity class.** In the database, we used the following classes to describe the activity: Seyfert class 1, 1.5, 1.8, 1.9, 2, LINER as Seyfert class 3, BL Lacertae, SB (staburst nuclei) and HII objects (spectra similar to HII regions). According to DR7 SDSS spectra, the galaxy SBS0953+574 is classified as DA star. According to the same DR7 data, two galaxies SBS1142+592A and SBS1146+596 are classified as QSOs. We should note that there are probably many more SBs and

HII objects among SBS galaxies that have narrow emission line spectra but for which detailed spectral information is not available yet. If there is not sufficient spectral information for the classification, a description of the available spectra as *e* (emission), *a* (absorption), or *e,a* (emission, absorption) is given.

**4. Measured parameters for SBS galaxies.**

**4.1. Observational material and images of SBS galaxies.** Central to this project was the availability of the digital sky survey (DSS) images obtained in support of the building of the *HST* Guide Star Catalogs (GSC, e.g. [40]). The GSC is based on Palomar Schmidt (POSS-II) plates for the northern sky. All plates were digitized at the Space Telescope Science Institute (STScI) using Perkin-Elmer PDS 2020G scanning microdensitometers with various modifications. The POSS-II plates were scanned at a resolution of 1.0" pixel$^{-1}$. These images have better resolution and fidelity as compared to the earlier POSS-I surveys which were used as additional observational material during the SBS survey.

Using positions of the SBS galaxies published by [37], 10' x 10' regions centered on each galaxy were extracted from the POSS-II J, POSS-II F and POSS-II IV-N images. Using these images, more accurate coordinates of the SBS galaxies were measured. In addition, the morphologies of the galaxies were determined as well as apparent magnitudes, major and minor diameters, position angles, and the apparent number of neighboring galaxies observed within a 50 kpc circle radius calculated from the published redshift.

**4.2. The coordinates.** In [37] the coordinates are presented with the accuracy of about ±1". The source of accurate coordinates for SBS galaxies is mainly [41]. For our study, the galaxy coordinates were measured from the POSS-II F plates in automatic mode according to the routine developed at STScI. Positions of stellar objects are typically located to better than 0.1" using either a 2D Gaussian fit or the pixels intensity-weighted moments. The actual positions of extended objects such as these are somewhat more poorly determined, because more complex morphologies add to the difficulty of locating the image centroids. For such objects and also for galaxies with multicomponent structures, we examined the images with the Aladin interactive software. We then checked and carefully measured the positions of the centroid centers of diffuse objects and those of the multiple components embedded in a common envelope. In these cases, positions may be uncertain to 1". The coordinates are in the *ICRS* system.

During the process of checking coordinates, we found that SBS1438+507A and B which according their coordinates in [37] form a very close pair is not real. It is in fact a single star-like galaxy, which is also confirmed by inspecting the SDSS DR7 image. According to this information we keep in our database only the SBS1438+507A. It is interesting to mention also that the object SBS1050+505B is an HII region in SBS1050+505A which is also the well known irregular galaxy Markarian 156 [29].

**4.3. The morphology.** In [37] the morphological information for the SBS galaxies was compiled from the literature when available. The current database presents a complete and homogeneous data for the morphologies of the SBS galaxies. The morphological classification of each galaxy was first carried out on the red image, and later checked using the blue image and sometimes also in the near-infrared image. We have carried out this classification by using gray-scale displays of the digitized images and by inspecting isophotal maps; the latter are especially useful in representing the large dynamic range of the images. In all cases when the sample galaxies SDSS DR7 images were available we checked our morphological classification using SDSS *g* and *r* images. We classified the SBS galaxies using the modified Hubble sequence (E-S0-Sa-Sb-Sc-Sd-Sm-Im) and intermediate cases, and the extension to blue compact dwarf galaxies (BCDs; [42]). In addition, 563 SBS galaxies could only be classified as objects with suspected spiral structure. 364 galaxies were classified as compact objects. 130 galaxies are close interacting or merging systems, classified as separate classes and most of them were studied in details in [43]. In five cases (SBS0921+519S, SBS1050+505B, SBS1050+573, SBS1551+601A and B) the SBS object is actually a giant HII complex in the large galaxy. The quality of our morphological classification system is discussed in [44].

**4.4. Apparent $J_{pg}$, $F_{pg}$ and $I_{pg}$ magnitudes.** In this database, we provide our measurements of the apparent, blue, red and near-infrared magnitudes for all SBS galaxies with improved accuracy and in homogeneous manner. The magnitudes of the galaxies were measured from the POSS-II photographic survey plates that are available at STScI and used for the construction of the GSC-II catalog [45]. The technique used for determining galaxy magnitudes was the same as described in [29]. The blue, red and near-infrared apparent magnitudes of the sample galaxies were measured from the *J*, *F* and *I* band images in a homogeneous way at the isophote corresponding to 3 times the background rms noise, which is approximately at the 25.3 mag arcsec$^{-2}$ level [29].

In the past, when creating databases for Markarian [29] and Kazarian [46] galaxies we compared our blue magnitude measurements with HYPERLEDA determinations. It was found that the mean absolute difference between both measurements was in the order of 0.4 magnitudes. A similar difference (0.41 ± 0.49, N = 1045) is estimated for this SBS galaxies sample too. We also compared our blue magnitudes with those blue photographic magnitudes of the former SBS catalog [37] (the accuracy of the determination is no worse that ±0.5 magnitudes). The mean absolute difference between both measurements is 0.11±0.13 (N=1676). A comparison of our $I_{pg}$ band magnitude measurements with HYPERLEDA *I*-band magnitude determinations was also conducted. The mean absolute difference between both measurements is 0.27±0.49 (N=927).

**4.5. Angular diameters, axial ratios, and position angles.** In [37] the angular sizes of the SBS galaxies were measured on the blue POSS-I prints. These diameters are eye estimates; hence they are not homogeneous and poorly accurate.

For this database major and minor angular diameters, axial ratios ($R=D_{minor}/D_{major}$) and position angles of SBS galaxies were measured in a homogeneous way from the blue images of the galaxies at the same (25.3 mag arcsec$^{-2}$) isophotal level that was used for the magnitude measurement. Position angles (P.A.) of the major axes are measured from the north (P.A.=0°) toward east between 0° and 180°.

We used data from HYPERLEDA to verify the agreement of our diameter system with the standard *D*(25) diameter system [e.g. 47] of HYPERLEDA and subsequently the agreement between ours and HYPERLEDA's axial ratio and position angle systems. Similar to the Markarian galaxies, for which the same technique for diameter measurement was used [29], our measured blue diameters are typically larger (1.4"±12.1", N=1087) than HYPERLEDA blue *D*(25) diameters. This is due to the deeper mean limiting galaxy surface brightness in our system.

A comparison of the SBS galaxies axial ratios in HYPERLEDA and our measurements shows no significant difference (0.0±0.17, N=1087). The same result was obtained also for Markarian galaxies [29].

A comparison of our P.A. blue measurements for the SBS galaxies with HYPERLEDA determinations reveal the same order of difference with that of the Markarian galaxies [29]. The mean difference between the HYPERLEDA's and our blue P.A.'s is 7.8°±7.8° N=948. Possible reasons for the large scatter between our and HYPERLEDA's P. A. are the same and discussed in [29].

**4.6. Counts of neighbor galaxies.** Counts of neighboring galaxies were done for all SBS galaxies which have determined redshifts *z*>0.005 by projecting a circle of 50 kpc radius around each galaxy. All galaxies detected within this circle were counted if their angular sizes differed from that of the SBS galaxy by no more than factor of 2 [e.g. 48], and wherever redshifts were available, have a velocity difference within ±800 km s$^{-1}$ [e.g. 49]. The counts of neighbor galaxies were checked in the 50 kpc circles extracted from all three $J_{pg}$, $F_{pg}$ and $I_{pg}$ -band images. There were 31 SBS objects closer than z=0.005 that were not used because of the difficulty in reliably determining associated objects over a wider field of view as random projections become more dominant.

## 5. The optical database.

To illustrate the form and content of the SBS galaxies database in Table 1 its first 10 objects are shown. The entire database for 1676 SBS galaxies is published in electronic form and is available in the VizieR Catalog Service via http://cdsarc.u-strasbg.fr/cgi-bin/VizieR?-source=VII/264. The database contains observational data for the 1676 SBS objects aligned in 17 columns, which are described below.

Column (1) – The SBS designation as it appears in [37].

Column (2) and (3) – Equatorial coordinates (ICRS).

Column (4) – The morphological description of the galaxy. The numerical coding used here for the morphological description of the galaxies is a slightly modified and simplified version of the morphological types T given in the RC3 catalog. The following codes were used: E=-5; E/S0=-3; S0=-2; S0/a=-1; S=0 (this is a class of suspected spiral galaxies); Sa=1; Sab=2; Sb=3; Sbc=4; Sc=5; Scd=6; Sd=7; Sdm=8; Sm=9; Im=10; Im/BCD=11; BCD/Im=12; BCD=13; Compact=14; Interacting system or Merger =15 and HII region in brighter galaxy =16. A bar is marked by "B".

Column (5) – Spectral classification according to [37].

Column (6) – Activity class, when available, or description of the spectra according to [37] and DR7 of SDSS. The various Seyfert classes are denoted by the symbols "Sy1", "Sy1.5 ", "Sy1.8", "Sy1.9", and "Sy2". The symbol "Sy3" refers to LINERs. Starburst nuclei are indicated by "SB" and HII galaxies are indicated by "HII". QSOs and BL Lac objects are also indicated.

Column (7) – Heliocentric redshifts when available. Redshifts are from [37] and also from SDSS DR7.

Columns (8), (9) and (10) – Apparent isophotal $J_{pg}$ (blue), $F_{pg}$ (red), and $I_{pg}$ (near-infrared) magnitudes.

Column (11) – Major $J_{pg}$ band diameter $D(J)$ in arcseconds.

Column (12) – Axial ratio in $J_{pg}$ band R$(J)$.

Column (13) – Position angle in $J_{pg}$ band P. A.$(J)$. It is measured from north (P.A.=0°) toward east between 0° and 180°. For round galaxies with axial ration $R(J)\approx1.00$ position angles were not measured.

Column (14) – Number of galaxies (N) detected within a 50 kpc projected radius. The mark "nd" (no data) in this column relates to SBS galaxies with redshifts smaller than 0.005 for which neighbor counts were not performed.

Columns (15), (16) and (17) – We provide near-infrared J, H, and K magnitudes for 1117 SBS galaxies. These data are from 2MASS determinations [50].

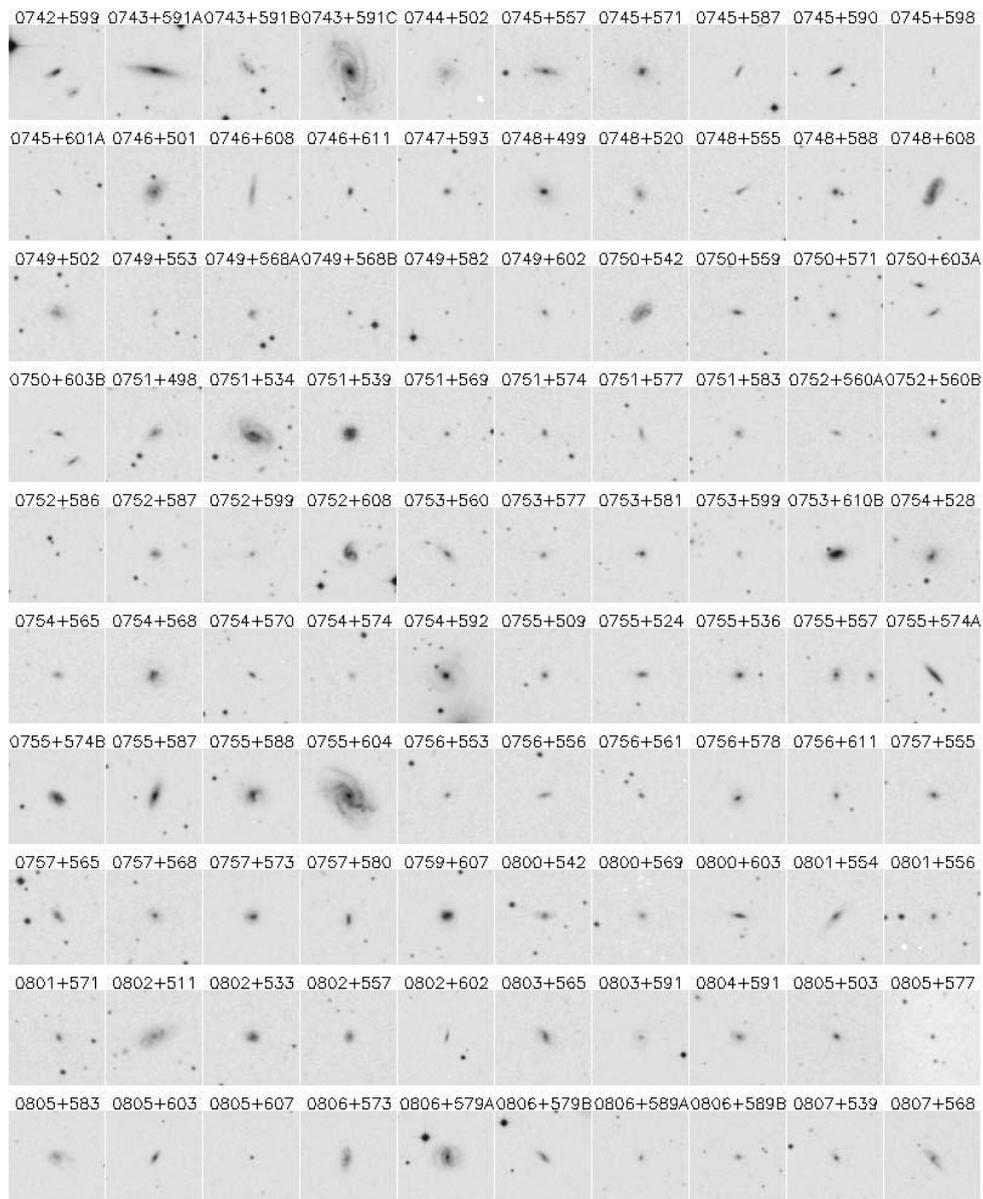

Fig. 1. 2' x 2' field POSS-II Jpg band images for the SBS galaxies. North is up and east is to the left.

In Figure 1 we show the first 100 2'x2' regions around each SBS galaxy from the digitized POSS-II $J_{pg}$ images. The complete atlas with each plate showing 100 SBS galaxies is available in the VizieR Catalog Service via http://cdsarc.u-strasbg.fr/cgi-bin/VizieR?-source=VII/264. On top of each image, the SBS galaxy designation is presented. The contrast of the images has been adjusted to provide the best subjective compromise in displaying the outer regions of the galaxies while preserving the structure of their inner regions.

**Acknowledgments.**

One of the authors, A.P., acknowledges the hospitality of the Space Telescope Science Institute (Baltimore, USA) during his stay as visiting scientist supported by the Director's Discretionary Research Fund. He also acknowledges the hospitality of the Institut d' Astrophysique de Paris (Paris, France) during his stay as visiting scientist supported by the Collaborative Bilateral Research Projects 2010 of the State committee of Science of the Republic of Armenia (SCS) and the French Centre National de la Recherche Scientifique (CNRS). This research has made use of the NASA/IPAC Extragalactic Database (NED), which is operated by the
Jet Propulsion Laboratory, California Institute of Technology, under contract with the National Aeronautics and Space Administration and HYPERLEDA (Leon-Meudon Extragalactic Database, http://cismbdm.univ-lion1,fr/~hyperleda). The Digital Sky Survey was produced at the Space Telescope Science Institute under U.S. Government grant NAG W-2166. The images of this survey are based on photographic data obtained using Oschin Schmidt Telescope on Palomar Observatory. The plates were processed into the present digital form with the permission of this institute. The Second Palomar Observatory Sky Survey (POSS-II) was made by the California Institute of Technology with funds from the National Science Foundation, and the National Aeronautics and Space administration, the National Geographic Society, the Sloan Foundation, the Samuel Oschin Foundation and the Eastman Kodak Corporation. The California Institute of Technology and Palomar Observatory operate the Oschin Schmidt Telescope. Funding for the SDSS and SDSS-II was provided by the Alfred P. Sloan Foundation, the Participating Institutes, the National Science Foundation, the U. S. Department of Energy, the national Aeronautics and Space Administration, the Japanese Monbukagakusho, the Max Planck Society and the Higher Education Founding Council for England. The SDSS was managed by the Astrophysical Research Consortium for the Participating Institutes. For the image processing the ADHOC software (www.astrsp-mrs.fr/index_lam.html) developed by Dr. Jacques Boulesteix (Marseille Observatory, France) was in intensive use.